# Fokker-Planck approach to non-Gaussian normal diffusion: Hierarchical dynamics for diffusing diffusivity

Sumiyoshi Abe [1-4]

[1] *Department of Physics, College of Information Science and Engineering, Huaqiao University, Xiamen 361021, China*

[2] *Institute of Physics, Kazan Federal University, Kazan 420008, Russia*

[3] *Department of Natural and Mathematical Sciences, Turin Polytechnic University in Tashkent, Tashkent 100095, Uzbekistan*

[4] *ESIEA, 9 Rue Vesale, Paris 75005, France*

**Abstract**   A theoretical framework is developed for the phenomenon of non-Gaussian normal diffusion that has experimentally been observed in several heterogeneous systems. From the Fokker-Planck equation with the dynamical structure with largely separated time scales, a set of three equations is derived for the fast degree of freedom, the slow degree of freedom and the coupling between these two hierarchies. It is shown that this approach consistently describes "diffusing diffusivity" and non-Gaussian normal diffusion.

## I. INTRODUCTION

Diffusion is characterized in such a way that a spatial scale $l$ and the duration time $t$ are related to each other as $l^2 \sim t^\nu$, where $l^2$ can be defined as the mean square displacement of a diffusing particle or the square of spatial extension of a probability distribution of an ensemble of such particles. $\nu$ stands for a positive diffusion exponent: the cases $\nu > 1$ and $\nu < 1$ are termed superdiffusion and subdiffusion, respectively. These are called the phenomena of anomalous diffusion [1] and form an integral part of various research areas in sciences. On the other hand, the case $\nu = 1$, i.e.,

$$l^2 = 2D_0 t \qquad (1)$$

with $D_0$ being a diffusion coefficient, describes normal diffusion that has been discussed in the context of Brownian motion [2]. It can be understood in terms of random walk of a particle in a homogeneous medium. The corresponding probability distribution of the walker's position is Gaussian,

$$p_G(x,t) = \frac{1}{\sqrt{4\pi D_0 t}} \exp\left[-\frac{x^2}{4D_0 t}\right] \qquad (-\infty < x < \infty,\ 0 \leq t) \qquad (2)$$

in the one-dimensional model, where the initial position is taken to be at the origin, and accordingly $l^2$ in Eq. (1) is identified with the variance of the displacement with respect to this distribution. Such a classical phenomenon is ubiquitously observed in nature.

About two decades ago, however, remarkable discoveries were experimentally made [3,4]. It has been manifested that normal diffusion occurs in heterogeneous colloidal systems near glass transition in spite of that fact that the probability distribution of particle displacement is non-Gaussian. In particular, the authors of Ref. [3] have pointed out that the observed non-Gaussian distribution is described by a sum of two Gaussians reasonably well.

Later, further results have been reported on experiments by use of colloidal beads on lipid bilayer tubes and in porous media created by entangled actin filaments [5], liposomes in nematic solutions of aligned actin filaments [6], and moisture-absorbing polymer films in the environment with controllable humidity [7]. In each of these systems, the probability distribution of particle displacement is again non-Gaussian and is well fitted by Laplacian

$$p(x,t) = \frac{1}{\sqrt{4D_0 t}} \exp\left(-\frac{|x|}{\sqrt{D_0 t}}\right), \tag{3}$$

which also gives rise to normal diffusion in Eq. (1). A characteristic feature is the presence of a cusp at $x=0$, in consistency with the experimental observations [5-7].

Recently, some attempts have been made in order to theoretically describe the phenomenon of non-Gaussian normal diffusion [8-10]. A basic idea is as follows. Because of heterogeneities of the media, the diffusion coefficients exhibit slow spatial, temporal or spatiotemporal variations. Thus, $D_0$ in Eq. (2) may not be a fixed constant but a value of realization of the random variable, $\mathcal{D}$, obeying a certain probability distribution $\Pi(D)$. This concept is called "diffusing diffusivity", and $\Pi(D)$ is referred to as the diffusivity distribution. That is, the Gaussian distribution in Eq. (2) should actually be regarded as a conditional probability distribution with a given value of $D$

$$p_G(x,t) \equiv p_G(x,t|D). \tag{4}$$

Then, the observed distribution of particle displacement is interpreted as the ensemble average

$$p(x,t) = \int_0^\infty dD \, p_G(x,t|D) \, \Pi(D). \tag{5}$$

Since $p_G(x,t|D)$ is assumed to be Gaussian in Eq. (2), use of the exponential distribution as $\Pi(D)$ [8]

$$\Pi(D) = \frac{1}{D_0} \exp\left(-\frac{D}{D_0}\right) \tag{6}$$

in Eq. (5) gives rise to Eq. (3). In this approach, $D_0$ in Eq. (3) is now the average of $\mathcal{D}$:

$$D_0 = \langle \mathcal{D} \rangle = \int_0^\infty dD \, D \, \Pi(D). \tag{7}$$

As noticed in Refs. [9,10], this corresponds to superstatistics [11,12]. The authors of Ref. [9] have generalized the above-mentioned formulation to the case when time $t$ in Eqs. (1)-(3) is replaced by $t^\nu$ (together with introduction of the generalized diffusivity, $D_*$) in order to cover anomalous diffusion. In the language of scaling, $p_G(x,t|D_0) = \left(1/\sqrt{D_0 t}\right) F_G\left(x/\sqrt{D_0 t}\right)$ is generalized to $p_G(x,t|D_*) = \left(1/\sqrt{D_* t^\nu}\right) F_G\left(x/\sqrt{D_* t^\nu}\right)$, where $F_G$ is the Gaussian scaling function given by $F_G(s) = \left(1/\sqrt{4\pi}\right)\exp\left(-s^2/4\right)$.

In the theoretical description mentioned above, the probability distribution of particle displacement for each value of realization of $\mathcal{D}$ is still assumed to be Gaussian in Eq. (2) leading to normal diffusion. In other words, the form of $\Pi(D)$ in Eq. (6) strictly depends on this point. It is known in mathematics [13] that a class of infinitely divisible probability distributions is wide. Therefore, experimental justifications of either $p_G(x,t|D)$ or $\Pi(D)$ in Eq. (6) are desired. Although the Gaussianity of $p_G(x,t|D)$ seems natural in view of the central limit theorem, strictly speaking it is not clear if particle displacements are independent and identically distributed. We will come back to this point in Sec. IV.

Here, we make an attempt to extract as much information as possible about the *hierarchical dynamics* underlying non-Gauusian normal diffusion from the property of the joint probability distribution,

$$P(x,D,t) = p(x,t|D)\,\Pi(D), \qquad (8)$$

by using the Fokker-Planck equation. We develop a method of adiabatic separation to explicitly describe the hierarchies.

This paper is organized as follows. In Sec. II, a general discussion is developed about the Fokker-Planck theory of two degrees of freedom that are characterized by largely separated time scales. There, separation of such a kinetic equation into the equations for the fast degree of freedom, slow degree of freedom, and the coupling between these two hierarchies is established. Then, in Sec. III, the theoretical framework presented in Sec. II is applied to the phenomenon of non-Gaussian normal diffusion. The roles of diffusing diffusivity are revealed. Finally, Sec. IV is devoted to concluding remarks, including additional experimental arguments about Eq. (6) and its generalization.

## II. HIERARCHICAL FOKKER-PLANCK EQUATION

To treat the particle displacement $X$ and diffusivity $\mathcal{D}$ as the random variables, let us define a two-tuple:

$$\mathbf{X} = \begin{pmatrix} X_1 \\ X_2 \end{pmatrix} = \begin{pmatrix} X \\ \Delta \end{pmatrix}. \tag{9}$$

In this notation, $\Delta$ denotes

$$\Delta = \ln\left(\mathcal{D}/\tilde{D}\right) \tag{10}$$

with $\tilde{D}$ being a positive constant that cancels the dimensionality of $\mathcal{D}$. This variable may be useful since $\mathcal{D}$ is nonnegative. Without loss of generality, $\tilde{D}$ can be set equal to unity,

$$\tilde{D} = 1, \tag{11}$$

and we will work in this unit.

Let $d^2\mathbf{x}\,\tilde{P}(\mathbf{x},t)$ be the probability of $\mathbf{X}$ being realized in $\mathbf{x} = (x_1, x_2)^T$ and $\mathbf{x} + d\mathbf{x} = (x_1 + dx_1, x_2 + dx_2)^T$ at time $t$. We require that the probability distribution, $\tilde{P}(\mathbf{x},t)$, obeys the Fokker-Planck equation in the following general form [14,15]:

$$\frac{\partial \tilde{P}}{\partial t} = -\sum_{i=1}^{2} \frac{\partial}{\partial x_i}\left(K_i \tilde{P}\right) + \sum_{i,j=1}^{2} \frac{\partial^2}{\partial x_i \partial x_j}\left(\sigma_{ij} \tilde{P}\right), \tag{12}$$

where $\sigma = (\sigma_{ij})$ is a positive semidefinite matrix, that is, symmetric with all of its eigenvalues being non-negative. We mention that a kinetic approach to description of the diffusivity distribution has been examined in Ref. [8]. However, in contrast to that approach, here we are considering the Fokker-Planck equation for the joint distribution.

It is convenient to change $x_2$ [i.e., the realization of $\Delta$ in Eq. (10)] to the original diffusivity variable, $D$, and therefore, $\partial/\partial x_2 = D\partial/\partial D$. Likewise, the probability distributions becomes $(1/D)\tilde{P}(x, \ln D, t)$, where $1/D$ is the Jacobian factor. This is the one to be identified with the joint probability distribution in Eq. (8). Thus, Eq. (12) is explicitly written as follows:

$$\frac{\partial P}{\partial t} = -\frac{\partial}{\partial x}\left(K_1 P\right) - D\frac{\partial}{\partial D}\left(K_2 P\right)$$
$$+ \frac{\partial^2}{\partial x^2}(\sigma_{11} P) + 2D\frac{\partial^2}{\partial x \partial D}(\sigma_{12} P) + D\frac{\partial}{\partial D}\left[D\frac{\partial}{\partial D}(\sigma_{22} P)\right], \tag{13}$$

where $K_i$'s and $\sigma_{ij}$'s are the functions of $(x, D, t)$, anew.

Now, we proceed into introduction of the hierarchical structure. $X$ and $\mathcal{D}$ are the fast and slow variables, respectively, and this justifies the decomposition in Eq. (8). The point is that the fast degree of freedom is significantly influenced by the slow degree of freedom, whereas the slow degree of freedom is independent of the fast degree of freedom. This may be analogous to the Born-Oppenheimer approximation [16] that is widely applied to the problems in quantum chemistry of molecules. Thus, we set

$$K_2 = K_2(D), \tag{14}$$

$$\sigma_{22} = \sigma_{22}(D). \tag{15}$$

Then, substituting Eq. (8) into Eq. (13) with Eqs. (14) and (15), we have

$$\begin{aligned}
\Pi(D)\frac{\partial p(x,t|D)}{\partial t} &= -\Pi(D)\frac{\partial}{\partial x}\left[K_1(x,D,t)p(x,t|D)\right] \\
&\quad -D\frac{\partial}{\partial D}\left[K_2(D)p(x,t|D)\Pi(D)\right] \\
&\quad +\Pi(D)\frac{\partial^2}{\partial x^2}\left[\sigma_{11}(x,D,t)p(x,t|D)\right] \\
&\quad +2D\frac{\partial}{\partial D}\left\{\Pi(D)\frac{\partial}{\partial x}\left[\sigma_{12}(x,D,t)p(x,t|D)\right]\right\} \\
&\quad +D\frac{\partial}{\partial D}\left\{D\frac{\partial}{\partial D}\left[\sigma_{22}(D)p(x,t|D)\Pi(D)\right]\right\}. \tag{16}
\end{aligned}$$

Based on the implementation of the hierarchical structure mentioned above, this equation is separated as follows:

$$\frac{\partial p(x,t|D)}{\partial t} = -\frac{\partial}{\partial x}\left[K_1(x,D,t)p(x,t|D)\right]$$

$$+\frac{\partial^2}{\partial x^2}\left[\sigma_{11}(x,D,t)p(x,t|D)\right] \qquad (17)$$

for the fast variable that is refereed to here as the conditional Fokker-Planck equation, and

$$-\frac{\partial}{\partial D}\left[K_2(D)p(x,t|D)\Pi(D)\right]$$

$$+2\frac{\partial}{\partial D}\left\{\Pi(D)\frac{\partial}{\partial x}\left[\sigma_{12}(x,D,t)p(x,t|D)\right]\right\}$$

$$+\frac{\partial}{\partial D}\left\{D\frac{\partial}{\partial D}\left[\sigma_{22}(D)p(x,t|D)\Pi(D)\right]\right\} = 0 \qquad (18)$$

for the slow variable and the coupling between the hierarchies through $\sigma_{12}$. Equation (18) immediately leads to

$$-K_2(D)p(x,t|D)\Pi(D) + 2\Pi(D)\frac{\partial}{\partial x}\left[\sigma_{12}(x,D,t)p(x,t|D)\right]$$

$$+D\frac{\partial}{\partial D}\left[\sigma_{22}(D)p(x,t|D)\Pi(D)\right] = c(x,t), \qquad (19)$$

where $c(x,t)$ is a certain function. Integrating Eq. (19) with respect to $x$ and using the normalization condition, $\int_{-\infty}^{\infty} dx\, p(x,t|D) = 1$, we have

$$-K_2(D)\Pi(D) + D\frac{\partial}{\partial D}\left[\sigma_{22}(D)\Pi(D)\right] = \int_{-\infty}^{\infty} dx\, c(x,t), \qquad (20)$$

provided that

$$\sigma_{12}(x, D, t) p(x, t|D) \to 0 \quad (x \to \pm\infty) \tag{21}$$

is naturally required [see Eq. (44) below]. Taking the integration of Eq. (20) over $D$ and using the notation in Eq. (7), we have

$$-\langle K_2(\mathcal{D}) \rangle - \langle \sigma_{22}(\mathcal{D}) \rangle = \int_0^\infty dD \int_{-\infty}^\infty dx\, c(x,t), \tag{22}$$

where the following condition has been imposed:

$$D\sigma_{22}(D) \Pi(D) \to 0 \quad (D \to 0, \infty). \tag{23}$$

This condition will in fact turn out to be satisfied [see Eq. (40) below]. Since the left-hand side in Eq. (22) is supposed to be finite [see Eqs. (40) and (43)], we have to set

$$c(x,t) = 0 \tag{24}$$

in order to avoid the divergence on the right-hand side in Eq. (22). Moreover, we can further separate Eq. (19) with Eq. (24). In fact, it can be rewritten as

$$\left\{ -K_2(D)\Pi(D) + D\frac{\partial}{\partial D}[\sigma_{22}(D)\Pi(D)] \right\} p(x,t|D)$$
$$+ \left\{ 2\frac{\partial}{\partial x}[\sigma_{12}(x,D,t)p(x,t|D)] + D\sigma_{22}(D)\frac{\partial p(x,t|D)}{\partial D} \right\} \Pi(D) = 0, \tag{25}$$

from which the following equations are derived:

$$-K_2(D)\Pi(D) + D\frac{\partial}{\partial D}\left[\sigma_{22}(D)\Pi(D)\right] = 0, \tag{26}$$

$$2\frac{\partial}{\partial x}\left[\sigma_{12}(x,D,t)p(x,t|D)\right] + D\sigma_{22}(D)\frac{\partial p(x,t|D)}{\partial D} = 0. \tag{27}$$

It is noted that this procedure is consistent since Eq. (26) is precisely equal to Eq. (20) with Eq. (24).

Consequently, we obtain three key equations: Eq. (17) as the conditional Fokker-Planck equation for the fast degree of freedom, Eq. (26) for the slow degree of freedom, and Eq. (27) for the coupling between these two hierarchies.

### III. NON-GAUSSIAN NORMAL DIFFUSION

We are now in a position to apply the framework developed in Sec. II to the phenomenon of non-Gaussian normal diffusion in view of diffusing diffusivity. Our purpose is to determine **K** and $\sigma$ in terms of the particle displacement and diffusivity. More specifically, we wish to clarify under what conditions the joint probability distribution in Eq. (8) with the Gaussian in Eq. (2) with Eq. (4) and the diffusivity distribution in Eq. (6) can be the solutions of Eqs. (17), (26), and (27).

First, let us analyze the conditional Fokker-Planck equation (17). Since the location of the peak of the Gaussian in Eq. (2) [with Eq. (4)] does not change in time, the drift term is absent:

$$K_1 = 0. \tag{28}$$

Substituting Eqs. (4) and (28) into Eq. (17), we have

$$-\frac{1}{2t}+\frac{x^2}{4Dt^2}=\sigma_{11}(x,D,t)\left(\frac{x^2}{4D^2t^2}-\frac{1}{2Dt}\right)$$
$$-\frac{x}{Dt}\frac{\partial\sigma_{11}(x,D,t)}{\partial x}+\frac{\partial^2\sigma_{11}(x,D,t)}{\partial x^2}. \qquad (29)$$

This equation has a solution

$$\sigma_{11}=D, \qquad (30)$$

as expected.

Second, doing a manipulation similar to the above for Eq. (26), we find that

$$K_2(D)=D\frac{d\sigma_{22}(D)}{dD}-\frac{D}{D_0}\sigma_{22}(D) \qquad (31)$$

is derived, using the diffusivity distribution in Eq. (6).

Finally, substituting Eq. (4) with Eq. (2) and a general value, $D$, into Eq. (27), we obtain

$$\frac{\partial\sigma_{12}(x,D,t)}{\partial x}-\frac{x}{2Dt}\sigma_{12}(x,D,t)+\left(\frac{x^2}{8Dt}-\frac{1}{4}\right)\sigma_{22}(D)=0. \qquad (32)$$

To analyze this equation, here we make an assumption that *the hierarchical structure is stationary*. That is,

$$\sigma_{12}=\sigma_{12}(x,D). \qquad (33)$$

Then, in order for Eq. (32) with Eq. (33) to hold at any time $t > 0$, the following equations must simultaneously be satisfied:

$$\frac{\partial \sigma_{12}(x, D)}{\partial x} - \frac{1}{4}\sigma_{22}(D) = 0, \tag{34}$$

$$-\frac{x}{2Dt}\sigma_{12}(x, D) + \frac{x^2}{8Dt}\sigma_{22}(D) = 0. \tag{35}$$

These two equations are consistent with each other, and the solution is found to be: $\sigma_{12}(x, D) = (x/4)\sigma_{22}(D)$. The odd parity of $\sigma_{12}(x, t)$ comes from the fact that the relevant part of the Fokker-Planck operator in Eq. (13) should have even parity, but this quantity is combined with the first-order derivative in $x$. This is consistent with the even-parity nature of the probability distributions in Eqs. (2) and (3). If the analysis would separately be made in the cases $x < 0$ and $x > 0$: $\partial x / \partial x = \partial(-x)/\partial(-x) = 1$. So, it may be useful to employ the 1-dimensional analog of the radial coordinate variable, $|x|$, and to define $\partial/\partial|x|$. Then, the solution is given by

$$\sigma_{12}(x, D) = \frac{|x|}{4}\sigma_{22}(D). \tag{36}$$

This indicates that the strength of the coupling between the hierarchies increases as $|x|$ becomes larger. It is also noted that Eq. (36) in fact satisfies the requirement in Eq. (21) with $p(x, t | D)$ being the Gaussian.

Still it is necessary to determine $K_2(D)$ and $\sigma_{22}(D)$. This can be done as follows. So far, we have found that the matrix, $\sigma$, has the form

$$\sigma = \begin{pmatrix} D & (|x|/4)\sigma_{22}(D) \\ (|x|/4)\sigma_{22}(D) & \sigma_{22}(D) \end{pmatrix}. \tag{37}$$

Since this $2 \times 2$ matrix has to be positive semidefinite, the conditions

$$\operatorname{tr}\sigma = D + \sigma_{22}(D) \geq 0, \tag{38}$$

$$\det\sigma = D\sigma_{22}(D) - \frac{x^2}{16}\sigma_{22}^2(D) \geq 0, \tag{39}$$

have to be satisfied. Equation (38) is reasonable. On the other hand, Eq. (39) seems problematic since it implies that $x$ and $D$ might not be independent but constrained, in general. This point can be overcome if $\sigma_{22}(D)$ has the form

$$\sigma_{22}(D) = \frac{16}{L^2}D, \tag{40}$$

where $L$ is a positive constant. Then, Eq. (39) gives rise to

$$|x| \leq L. \tag{41}$$

Therefore, if $L$ is large enough, then the problematic point mentioned above will practically be resolved. A physical significance of $L$ is as follows. From Eqs. (2) and (3), Eq. (41) implies

$$\sqrt{Dt} \ll L, \tag{42}$$

which puts a constraint on the upper bound of the elapsed time, $t$, given the values of $D$ and $L$. In general, this is a complicated issue since $D$ is distributed according to $\Pi(D)$. One possible way to overcome it is to set a cutoff, $D_{max}$, as in Ref. [8]. This is natural from the experimental viewpoint, and, in this way, the upper bound of the elapsed time in measurement can be evaluated. We also note that Eq. (40) justifies the condition in Eq. (23). Finally, substituting Eq. (40) into Eq. (31), we have

$$K_2(D) = \frac{16}{L^2} D \left(1 - \frac{D}{D_0}\right). \qquad (43)$$

In addition, we also have

$$\sigma_{12}(x, D) = \frac{4}{L^2} |x| D, \qquad (44)$$

from Eqs. (36) and (40). Equations (40) and (43) assure that the left-hand side in Eq. (22) is finite, as supposed.

To summarize, we have determined all of the quantities responsible for non-Gaussian normal diffusion: Eqs. (28) and (30) for the fast degree of freedom, Eqs. (40) and (43) for the slow degree of freedom, and Eq. (44) for the stationary coupling between these two hierarchies.

## IV. CONCLUDING REMARKS

We have formulated a theoretical framework for describing an exotic phenomenon of non-Gaussian normal diffusion based on the stochastic process and the Fokker-Planck theory, in which both the particle displacement and diffusivity are treated as the random variables. Taking advantage of the large separation of time scales in the dynamics, we have developed a discussion that enables us to reveal the hierarchical structure underlying the phenomenon. We have determined all of the system-specific quantities in consistency with the joint probability distribution of the particle displacement and diffusivity proposed in the literature.

This paper is a generalization of the one presented in Ref. [17], in which only the stationary solution of the Fokker-Planck equation with a hierarchical structure is discussed.

The form of the joint probability distribution in Eq. (8) with Eqs. (4) and (6) is, as mentioned in Sec. I, a basic premise, and it is desirable to directly measure either the conditional Gaussian distribution in Eqs. (2) and (4) or the diffusivity distribution in Eq. (6), experimentally. Actually, the exponential diffusivity distribution has been inferred in the system of RNA-protein particles in the cellular cytoplasm [18], in spite of the fact that the diffusion property observed there has been subdiffusion, not normal diffusion. A description of such a probability distribution by use of the maximum entropy method is discussed in Ref. [19].

We wish to mention an interesting discussion presented in recent works in Refs. [20,21] about exponential distributions such as the Laplacian in Eq. (3). It is based on the large deviation theory for continuous time random walks [22] with subordinations, where the number of a walker's jumps is random [1], corresponding to diffusing diffusivity.

Our final comment is on the claim made in Ref. [8] that the diffusivity distribution in Eq. (6) may have a power-law correction, $\Pi(D) = N D^\alpha \exp(-D/D_0)$, where $\alpha$ is a positive constant and $N = 1/\left[D_0^{\alpha+1} \Gamma(\alpha+1)\right]$ is the normalization constant. In this case, Eq. (31) changes as, $K_2(D) = D d\sigma_{22}(D)/dD - (D/D_0 - \alpha)\sigma_{22}(D)$. A discussion similar to that in Sec. III can be made also in such a case.

**ACKNOWLEDGMENTS**

This work has been supported in part by a grant from National Natural Science Foundation of China (No. 11775084), the Program of Fujian Province, and the Program of Competitive Growth of Kazan Federal University from the Ministry of Education and Science of the Russian Federation, all of which the author gratefully acknowledges.

---